\documentclass[a4paper,11pt,article,oneside]{memoir}
\usepackage{ecis2024}
\usepackage{tikz}
\usepackage{enumitem}
\usepackage{amsmath}
\usepackage{amssymb}
\usepackage{mathtools}
\usepackage{amsthm}
\usepackage{csquotes}
\usepackage{subcaption}
\usepackage{threeparttable}
\usepackage{float}
\usepackage[capitalize,noabbrev]{cleveref}

\newcolumntype{P}[1]{>{\centering\arraybackslash}p{#1}}
\newcolumntype{R}[1]{>{\raggedleft\let\newline\\\arraybackslash\hspace{0pt}}m{#1}}

\setlist[itemize]{align=parleft,left=0.5em..1.5em}

\newlist{rquestion}{enumerate}{2}
\setlist[rquestion,1]{before=\itshape,font=\itshape, ref=RQ\arabic*,}

\newlist{req}{enumerate}{2}
\setlist[req,1]{label=RQ \arabic*:,ref= \arabic*, leftmargin=*}

\newlist{sa}{enumerate}{2}
\setlist[sa,1]{before=\itshape,font=\itshape, label=Sample: ,ref=Sample \arabic*, leftmargin=*}
\newlist{trp}{enumerate}{2}
\setlist[trp,1]{before=\itshape,font=\itshape, label=Training Phase: ,ref=Training Phase \arabic*, leftmargin=*}
\newlist{tep}{enumerate}{2}
\setlist[tep,1]{before=\itshape,font=\itshape, label=Test Phase: ,ref=Test Phase \arabic*, leftmargin=*}
\newlist{ro}{enumerate}{2}
\setlist[ro,1]{before=\itshape,font=\itshape, label=Round: ,ref=Round \arabic*, leftmargin=*}

\newlist{hyp}{enumerate}{2}
\setlist[hyp,1]{before=\itshape,font=\itshape, label=Hypothesis \arabic*:,ref= \arabic*, leftmargin=*}
\setlist[hyp,2]{before=\itshape,font=\itshape, label=Hypothesis \arabic*:,ref= \arabic*, leftmargin=*}

\maintitle{Transferring Domain Knowledge with (X)AI-Based Learning Systems} 
\shorttitle{(X)AI-Based Learning Systems} 
\category{Completed Research Paper} 

\authors{
Philipp Spitzer, Karlsruhe Institute of Technology, Karlsruhe, Germany, philipp.spitzer@kit.edu

Niklas K{\"u}hl, University of Bayreuth, Bayreuth, Germany, kuehl@uni-bayreuth.de

Marc Goutier, Technical University of Darmstadt, Darmstadt, Germany, goutier@ise.tu-darmstadt.de

Manuel Kaschura, Karlsruhe Institute of Technology, Karlsruhe, Germany, office@ksri.kit.edu

Gerhard Satzger, Karlsruhe Institute of Technology, Karlsruhe, Germany, gerhard.satzger@kit.edu
}

\abstracttext{
In numerous high-stakes domains, training novices via conventional learning systems does not suffice. To impart tacit knowledge, experts' hands-on guidance is imperative. However, training novices by experts is costly and time-consuming, increasing the need for alternatives. Explainable artificial intelligence (XAI) has conventionally been used to make black-box artificial intelligence systems interpretable. In this work, we utilize XAI as an alternative: An (X)AI system is trained on experts‘ past decisions and is then employed to teach novices by providing examples coupled with explanations. In a study with 249 participants, we measure the effectiveness of such an approach for a classification task. We show that (X)AI-based learning systems are able to induce learning in novices and that their cognitive styles moderate learning. Thus, we take the first steps to reveal the impact of XAI on human learning and point AI developers to future options to tailor the design of (X)AI-based learning systems.
}


\begin{document}



\chapter{Introduction}
\label{sec: Introduction}

In an ever-changing world, organizations face the challenge of adapting to new market demands and standards \citep{markus2004technochange}. Through organizational learning, humans acquire knowledge and improve skills on the job to adapt to changes in such an environment, allowing organizations to evolve and remain competitive in the marketplace
\citep{levy2011knowledge, eirich2022life}. Learning systems are crucial to support this process of acquiring new skills \citep{argote2014knowledge}. Conventional learning systems like e-learning systems \citep{wong2011effects} or intelligent tutoring systems \citep{kochmar2022automated} have been proven to promote learning among employees within organizations. Yet, conventional learning systems are used in specific contexts \citep{derouin2005learning} and can be costly to develop \citep{serban2020large}. Thus, it is inevitable to understand how different approaches aid learning, especially in high-stakes domains where decision-making is critical.

The advent of artificial intelligence (AI) promotes the development of learning systems that are scalable and generalizable to overcome the drawbacks of conventional systems. For example, intelligent tutoring systems use AI to automate domain knowledge representation, which can be a time-consuming task \citep{gross2015learning}. Machine teaching represents another type of learning approach that uses AI to teach novices (inexperienced humans) tasks by providing a set of examples \citep{zhu2018overview}. Recent advances in this area use explainable AI (XAI). Traditionally, XAI is used to make AI advice more transparent and interpretable \citep{adadi2018peeking, van2022explainable}. Research in machine teaching uses XAI to provide additional explanations on top of these examples \citep{goyal2019counterfactual, spitzer2022training, vandenhende2022making}. Their results suggest promising applications for learning systems. However, these approaches to teaching humans have not been evaluated in user studies. Yet, (X)AI-based learning systems open up new ways of learning: Instead of automating specific tasks based on the data they are trained on, (X)AI-based learning systems are used to transfer the expert knowledge encoded in the data that represents experts' past decision behaviors. This way, explicit and tacit knowledge can be conveyed to novices \citep{sanzogni2017artificial}. Imagine a novice quality inspector that is new to identifying scrap characteristics on parts. Instead of being trained by an expert quality inspector manually showing relevant areas on parts, an (X)AI-based learning system can showcase relevant parts and features. To fully understand how such (X)AI-based learning systems aid human learning, it is necessary to advance the current research.



As (X)AI-based learning systems constitute an alternative approach to conventional learning, it is crucial to understand the driving factors in this process. In this socio-technical system \citep{bostrom1977mis} of human and AI, previous research points out that humans prefer different kinds of explanation modalities \citep{szymanski2021visual, hernandez2020explaining}. For instance, humans who prefer visual information, might prefer visual explanations (i.e., visualization maps) whereas humans who process textual information better might prefer textual explanations (i.e., natural language explanations). Prior research shows that cognitive styles are relevant factors in learning \citep{mayer2003three} and important for the XAI design \citep{schneider2019personalized}. While novices can have different cognitive styles \citep{kirby1988verbal} and the task they are learning can vary in its type (e.g., visual versus textual), it is crucial to investigate how novices' cognitive styles impact the learning performance when taught by (X)AI-based learning systems.

With the need in research and practice to scrutinize how novices can learn and build new knowledge in various domains, we take a first step in this direction and present the following research questions:

\begin{rquestion}[leftmargin=1.1cm, labelindent=0pt, labelwidth=0em, label=\textbf{RQ\arabic*}:, ref=RQ\arabic*]
    \item To what extent does explainable AI (XAI) facilitate the learning of novices?\label{RQ1}
\end{rquestion}

\begin{rquestion}[resume, leftmargin=1.1cm, labelindent=0pt, labelwidth=0em, label=\textbf{RQ\arabic*}:, ref=RQ\arabic*]
    \item How does the cognitive style impact the learning of novices when taught through explainable AI? \label{RQ2}
\end{rquestion}

To answer both research questions, we conduct a study with 249 participants. In the study, participants are tasked to classify mammography images. Through a between-subjects design, we establish two conditions in which we provide examples of mammography images without and with explanations generated by XAI. The (X)AI system is trained on images and annotations of medical experts, incorporating their expert knowledge through information on their past decision-making. We then examine how the provision of explanations in the form of visual highlights through this (X)AI-based learning system can foster novices' learning performances. To answer \ref{RQ1}, we examine how the provision of explanations through an (X)AI-based learning system can foster the learning performance \citep{perlich2003tree, kuhl2022human}.
Additionally, we collect data on participants' cognitive styles to address \ref{RQ2}. 

Our results reveal a higher learning performance of novices when provided with explanations by an XAI. Aligning with related research, our findings show that novices with a visual cognitive style achieve higher performance than non-visual novices in this visual classification task. We find that the visual cognitive style mediates the effect of example-based learning on novices' learning performance. We thereby add empirical insights to the rising field of learning systems in IS and take the first steps to understand the potential of XAI in learning. More precisely, in this work, we contribute to a better and more integrated understanding of how (X)AI-based learning systems can facilitate knowledge transfer.


\chapter{Related Work}
\label{sec: Related Work}

In the last years, there has been increased research into using AI systems not only to support humans in their interaction with AI to take over auxiliary tasks \citep{bullock2020mapping} but also to improve the training of novices to teach them new tasks \citep{cakmak2012algorithmic, edwards2018teacher}. 

The focus of recent research in human-AI collaboration is on XAI: The AI system is providing additional explanations to the novice \citep{xu2019explainable} to increase their ability to judge the AI's predictions \citep{adadi2018peeking}.
Recent studies explore XAI use to avoid overreliance, i.e., to not blindly follow AI advice \citep{ schemmer2022influence}. This is crucial as humans who blindly rely on the AI might oversee incorrect AI advice which leads to ineffective human-AI collaborations. \citet{schemmer2022meta} introduce the conceptualization of appropriate reliance to measure the reliance behavior of humans in human-AI collaborations. \citet{Dellermann2019} argue that in such collaborations, humans and AI  can learn from each other. In the empirical study of \citet{chen2023machine}, the authors explore the development of humans' understanding of the task domain in the context of collaborative settings with an AI. 

With the advance of research on XAI, only a few studies investigate the potential of explanations to foster novices' understanding of the underlying domain \citep{schemmer2023towards, jussupow2021augmenting, bauer2023expl}. Instead of using XAI to make human-AI decision-making more interpretable to humans, XAI is used to convey expert knowledge. 
Machine teaching~\citep{zhu2018overview} presents one field in recent research scrutinizing teaching concepts of learning systems by utilizing AI. The underlying concept is to select the optimal teaching set, assuming that the teacher is aware of the decision boundaries \citep{zhu2018overview}. A teaching set comprises instances of the dataset that are presented to a learner. Research in this area is concerned with how to select a teaching set to achieve the best learning performance \citep{singla2014near}. An example is the study of \citet{johns2015becoming} in which the authors analyze several AI-based teaching strategies by choosing the optimal teaching set online. Furthermore, \citet{su2017interpretable} investigate how additional explanations via feature feedback increase novices' learning in an example-based learning systems. They show that explanations improve the understanding of underlying concepts and accelerate the learning progress. 

Consequently, recent studies advance the XAI techniques to generate counterfactual explanations. Counterfactual explanations are examples that represent instances of a different class than the current instance, where the features change minimally compared to the current instance. \citet{goyal2019counterfactual} introduce counterfactual explanations based on discriminative regions (areas in the images that are important for the AI to classify them properly). \citet{goyal2019counterfactual} argue that such explanations represent a teacher who \enquote{explain[s] why something is a particular object and why it's not some other object} \citep[p. 2377]{goyal2019counterfactual}. In a different study, \citet{vandenhende2022making} revise such counterfactual explanations based on discriminative regions. They ensure that the regions indicated in the original image and in the counterfactual image have the same semantic meaning.

Recent research on human-AI collaboration focuses on the human user in the design and development of XAI methods \citep{zhu2018explainable, liao2021human}. \citet{miller2019explanation} argues that humans' processing of explanations is inherently unique and that humans' cognitive styles affect the information process. In general, it is known that humans \enquote{process the same information in different ways} \citep[p. 267]{riding1993individual}. Moreover, based on a computer-supported method, \citet{riding1993individual} investigate different cognitive styles of humans. The authors distinguish cognitive styles along two dimensions: Whereas the Wholist-Analytic style describes the organization of information, the Verbal-Imagery style focuses on the representation of information: While verbals (humans with a preferred verbal style) tend to prefer textual information, visuals (humans with a preferred visual style) learn better from images \citep{riding1993individual}. Hence, in terms of thinking, visuals tend to use mental pictures while verbals use words \citep{riding1991cognitive}. Empirical studies, for instance in the domain of recommender systems, investigate how different representation styles of explanations are perceived by humans \citep{hernandez2020explaining}. In \citet{riefle2022influence}, the authors examine how humans' cognitive styles impact their comprehension of explanations generated through XAI.

These studies show (especially from a technical point of view) that such explanations can aid visual examples to differentiate certain visual patterns from each other and offer a promising approach to investigate the role of XAI in learning empirically. Thus, in this article, we empirically explore the potential of (X)AI-based learning systems to convey expert knowledge.

\chapter{Theoretical Development}
\label{sec: tdev}

The growth in the use of explanations to reveal the rationale behind AI predictions has led to an increase in research studying the impact of explanations on decision-makers' behavior \citep{de2022perils, schemmer2022meta, leichtmann2023explainable}. With the gap in research to thoroughly analyze the effect of XAI on learning, we aim in this work to shed light on the capabilities of (X)AI-based learning systems. We do this by conducting an online study and empirically investigating how XAI affects novices' learning performance for a visual classification task in an example-based learning setting. 
In general, example-based learning describes the process of providing examples to novices to enhance their learning \citep{van2010example}. This improvement in learning can lead to the development of novices' mental models of a task's underlying concept \citep{michael2004mental}. 
Moreover, \citet{cai2019effects} investigate how explanations in an example-based setting impact the user's mental model on the AI system. Their findings show that various kinds of explanations affect the perception of the AI system in different ways. Furthermore, research in cognitive science investigates the effect of explanations on learning. \citet{williams2010does} show that explanations can lead to an increase in learning to classify instances into the correct category. In our study, we explore how explanations generated by XAI affect participants' learning performance for a visual classification task. Thus, providing examples with visual explanations throughout the study will likely improve the classification ability of participants. Hence, we hypothesize:
\begin{hyp}[resume, wide, leftmargin=0cm, labelindent=0pt, labelwidth=0em]
\item Teaching novices with visual explanations generated by explainable AI (XAI) will lead to a higher learning performance for a visual classification task than a baseline in which sole examples are provided.
\label{hyp1}
\end{hyp}

On top of that, previous research observes the learning time of novices to account for the effectiveness of learning systems \citep{rasheed2021learning}. This time indicates how long novices need to process all information provided during training. Accordingly, we investigate the time spent interacting with the learning system. More precisely, we measure the time participants spend looking at the training samples. Previous research in HCI finds that additional information can lead to a higher cognitive load \citep{hudon2021explainable, herm2023impact}. In our study, we compare participants' learning for two conditions: In the control condition, participants learn based on examples, while in the XAI condition, participants are provided additional explanations generated through XAI. As we present additional information in the XAI condition, this information provision will likely result in longer learning times. Thus, we hypothesize:

\begin{hyp}[resume, wide, leftmargin=0cm, labelindent=0pt, labelwidth=0em]
\item Teaching novices with visual explanations generated by explainable AI (XAI) will lead to longer learning times compared to a baseline in which sole examples are provided.
\label{hyp2}
\end{hyp}

As related research on humans' cognitive styles shows, the learning performance of humans can depend on their cognitive styles \citep{riefle2022influence}. Expanding upon this premise, we propose that within visually intricate task domains, like the classification of mammography images, the cognitive styles of individuals can influence their learning performance. \citet{riding1993individual} introduce the cognitive style dimension along which humans process verbal and visual information differently. Thus, humans with a visual cognitive style who process visual information better might use the information in a visual classification task more effectively. With research in HCI investigating various explanation modalities  \citep{robbemond2022understanding}, it is crucial to understand how explanations of these specific modalities affect their learning performance.
Hence, we take the first steps to a more integrated understanding of the influence of human factors on the learning performance of novices. Visual explanations displaying additional information on top of the examples provided will most likely affect humans' performance in learning and understanding of a new concept, depending on their cognitive style. Consequently, we assume that novices' cognitive style will affect the learning performances and learning times throughout the study. As participants with a visual cognitive style process visual information better, we assume that they will decrease their learning times more compared to non-visual participants. Thus, we state the following:

\begin{hyp}[resume, wide, leftmargin=0cm, labelindent=0pt, labelwidth=0em]
\item Novices with a visual cognitive style achieve higher learning performances compared to participants with a non-visual cognitive style in a visual classification task when provided with visual explanations generated by explainable AI (XAI).
\label{hyp3}
\end{hyp}

\begin{hyp}[resume, wide, leftmargin=0cm, labelindent=0pt, labelwidth=0em]
\item Novices with a visual cognitive style will decrease their learning times compared to participants with a non-visual cognitive style in a visual classification task when provided with visual explanations generated by explainable AI (XAI).
\label{hyp4}
\end{hyp}

\begin{figure}[htbp!]
    \centering{\includegraphics[scale=0.64]{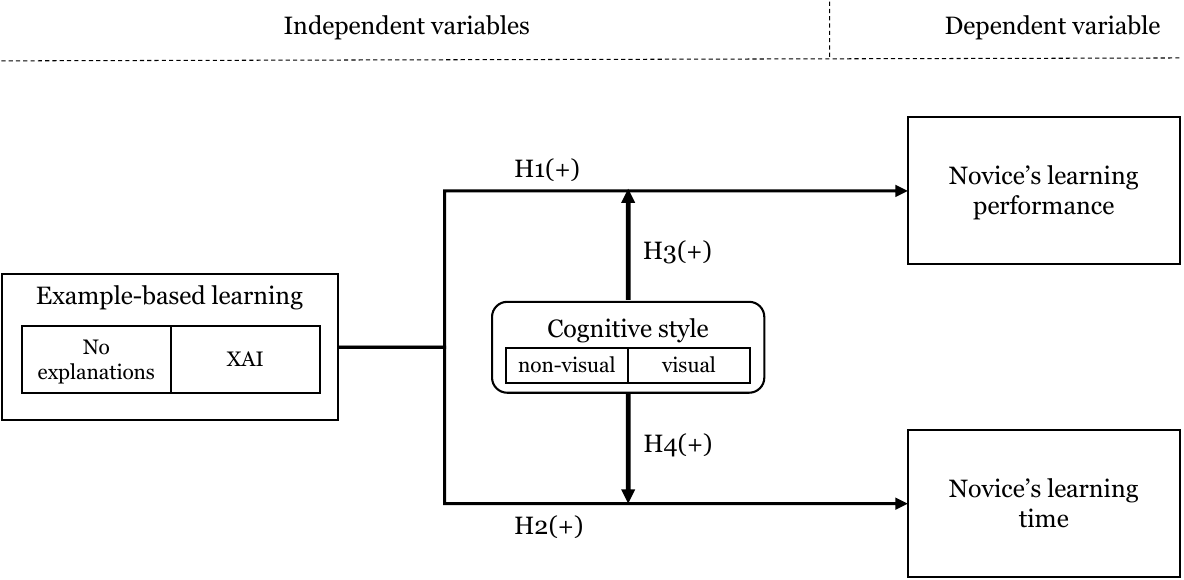}}
 \captionsetup{justification=raggedright,singlelinecheck=false}
    \caption{Research model.}
    \label{research_model}       
\end{figure}

\Cref{research_model} shows our overall research model and highlights the interrelation between independent and dependent variables. As can be seen in the figure and in alignment with \ref{RQ1} and \ref{RQ2}, we explore how XAI and different human factors influence novices' learning performance for a visual classification task.

\chapter{Study Design}
\label{sec: stdes}

To investigate the effect of XAI on teaching novices, we set up a user study with a between-subjects design. According to our research model \Cref{research_model}, we measure the learning performance and learning time of novices and examine whether novices' cognitive styles affect their learning. 

\section{Data selection}
\label{data_selection}
\begin{figure}[htbp!]

\centering
\begin{subfigure}{.3\textwidth}

  \centering
  \includegraphics[width=.4\linewidth]{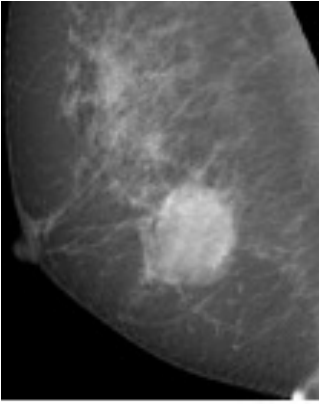}
    \captionsetup{justification=centering}

  \caption{Original Image.}
\end{subfigure}%
\begin{subfigure}{.3\textwidth}
  \centering
  \includegraphics[width=.4\linewidth]{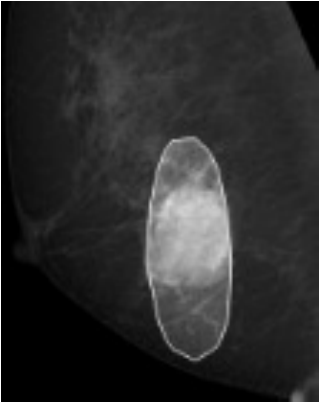}
    \captionsetup{justification=centering}

  \caption{Image with Explanation.}
  \centering
\end{subfigure}

    \centering
\captionsetup{justification=raggedright,singlelinecheck=false}
\caption{Example of an image without visual explanation (left) and the same image with corresponding explanation (right). The explanation highlights the relevant region with cancer cells.}
\label{example}

\end{figure}

The task of the study is to classify mammography images of human breasts. We use this classification task as it represents a real-world task that requires a certain level of expertise. We choose the well-established large-scale dataset of \citet{nguyen2022vindr} from \citet{goldberger2000physiobank}. We use two classes of the dataset participants have to distinguish: \textit{Cancer (positive)} or \textit{no cancer (negative)}. An example of a positive image is shown in \Cref{example} $a)$ on the left.

The dataset contains 5,000 four-view mammography examinations, complete with breast-level assessments and detection annotations. This information is used to train the (X)AI system. Each of these examinations underwent two independent readings to resolve disagreements between the reviewers. After consultation with a medical field expert, we only include images in the study with visual patterns that can be found through a thorough investigation of the respective image. Overall, we sample 60 images with cancer (positive) and 60 images without cancer (negative). From this pool of images, an equal amount of positive and negative images is shown to participants.

\section{Recruitment}
We recruit the participants for our study through the Prolific.co platform \citep{palan2018prolific}. Previous research shows that this platform presents a reliable source for research data \citep{peer2017beyond, palan2018prolific}. In total, we recruit 249 participants. The study was reviewed and approved by an Institutional Review Board before being conducted. 

We ensure that the sample is balanced in gender, and we only select participants living in the US or UK. By applying a filter for being fluent in the English language, we additionally ensure that all instructions and the entire study are well comprehended.
Participants who meet the stated eligibility criteria and complete the study's requirements receive a payment of 2.20 pounds. On top of that, participants are incentivized to thoroughly screen the provided data and classify the images correctly (4 cents per correct classification).

\section{Development of the XAI system}

To convey expert knowledge to novices, we develop an (X)AI-based learning system based on past expert decisions. In more detail, we train an AI from labels on mammography images generated by experts. Even though these annotations can still include biases and inconsistencies of the experts \citep{lebovitz2021ai, muller2021designing} we argue to use these labels for the development of the AI-based learning system. Firstly, the dataset used in this study provides multiple expert labels for each instance, minimizing the risk of these human biases and inconsistencies \citep{nguyen2022vindr}. Secondly, in a real-world setting, AI-based learning systems based on expert labels represent an alternative to human-based teaching. Gathering actual ground truth labels can be costly and resource-intensive. As AI-based learning systems intend to employ new knowledge continuously over time—based on organizations’ expert knowledge conveyed in their employees’ heads— expert labels are a trade-off for potential inconsistencies with the actual ground truth. Thus, in our study, we train the AI based on such expert labels. The visual explanations are generated using GradCam as XAI method \citep{selvaraju2017grad}. The GradCam method highlights regions in the image that are important for the AI to come up with its prediction. This methodology of visually emphasizing the important regions in an image is similar to the actual annotations that the medical experts provide. Similar to related works in this domain (i.e.,  \citet{cabitza2023rams}), we use these explanations to facilitate novices' learning of the task by showing these visual explanations. To do so, we train a convolutional neural network (ResNet-50) on the dataset \citep{he2016deep} and apply the GradCam method. In order not to hide relevant information on the image, we use a threshold to indicate only the most relevant parts of the visual highlights (see \Cref{example}). These areas visually highlight important parts of the image that the AI model uses for its prediction. Thus, they represent visual explanations that are displayed in the examples provided to novices. Based on this method, the (X)AI-based learning system can convey expert knowledge based on the data it is trained on.
The explanations are generated for each example that is shown to participants in the XAI condition. Participants can zoom in on the images to identify the relevant patterns indicated in the visual explanations.

\section{Study design}

The study begins with an introduction to the topic, which describes the task they will be asked to complete. 
We design the study through a between-subjects design with two conditions of different training phases. In the first condition (control condition), participants are provided with examples of positive and negative images. In the second condition (XAI condition), participants are provided with additional XAI generated explanations.
Through a pre-test with 15 participants, we evaluate the design decisions of the (X)AI-based learning system. This pre-test reveals that showing participants more than one image per class per round can lead to cognitive overload \citep{kirsh2000few, kuhl2022human}. On top of that, we derive the following design decisions:
Participants in the XAI condition are not informed that the explanations provided are generated by XAI. We make this design choice because past research shows that the awareness that the collaborator is an AI can impact their behavior \citep{dietvorst2015algorithm}.

Before starting the actual task in the study, participants are pre-evaluated on six randomly distributed images. This is done to assess participants' prior knowledge of the task before beginning the learning phase. To ensure participants' attention, as suggested by \citet{abbey2017attention}, they are asked to choose the task of the study after it is introduced.

In the next phase of the study, participants begin the task. Inspired by \citet{kuhl2022human}, the study is set up in iterative rounds, ten in total.
In each round, participants are first shown a training sample of each class and the corresponding ground truth label, which describes the true class of the image. We ensured the AI reached a sufficient performance level and only select instances for which the AI prediction is correct to explore the potential impact on if and how (X)AI can facilitate novices’ learning. In the XAI condition, participants also see explanations. We ensure that these explanations correspond to the annotations of the medical experts given in the dataset. We do this as we want to study the potential impact of such explanations on novices' learning. After the training phase in each round, participants are tested on four randomly selected samples. We ensure that both classes (positive and negative) are evenly distributed so that, on average, each participant sees the same number of positive and negative images. During the testing phase, the ground truth label is not revealed. All testing samples of a round are displayed next to each other on the same page. The study ends with questionnaires on participants' cognitive styles, their cognitive load, and their demographics.

In the testing phase of each round, we measure the learning performance of participants to derive the overall learning curve \citep{perlich2003tree}. We measure the learning performance by utilizing accuracy as a performance metric.
Additionally, we measure participants' learning time in each of the ten rounds.

We also collect data on participants' cognitive styles. We use the validated items of \citet{kirby1988verbal} through a post-task questionnaire and follow thereby the protocol of \citet{riefle2022influence}. The items of the cognitive style questionnaire are randomized as suggested by \citet{kirby1988verbal}. They are measured on a five-point Likert scale (``strongly disagree'', ``disagree'', ``neither disagree nor agree'', ``agree'', or ``strongly agree'') and are displayed to participants in random order.

\chapter{Results}
\label{sec: results}

In this section, we posit our results to answer \ref{RQ1} and \ref{RQ2}. To address \ref{RQ1}, we conduct thorough statistical analyses and present the findings in \Cref{res_xai}. We address \ref{RQ2} by performing a moderation analysis which we present in \Cref{res_cogstyl}. We corroborate our findings with qualitative results by analyzing the study's open-text questions.

Of the participants, 249 pass the attention checks and finish the study in accordance with the guidelines. The pre-evaluation serves as a means to confirm that participants possess no prior familiarity with the classification of mammography cancer images and can be regarded as novices. 
Among the 249 participants, 124 are female, 124 are male, and 1 is diverse. The average age of participants is 35.73 years ($std = 12.82$). Overall, it takes the participants on average 13 minutes and 50 seconds to complete the study. In total, 127 participants take part in the XAI condition and 122 participants in the control condition. 

\section{Impact of XAI on novices' learning}
\label{res_xai}

\begin{figure}[htbp!]

    \centering{\includegraphics[width=1.0\textwidth]{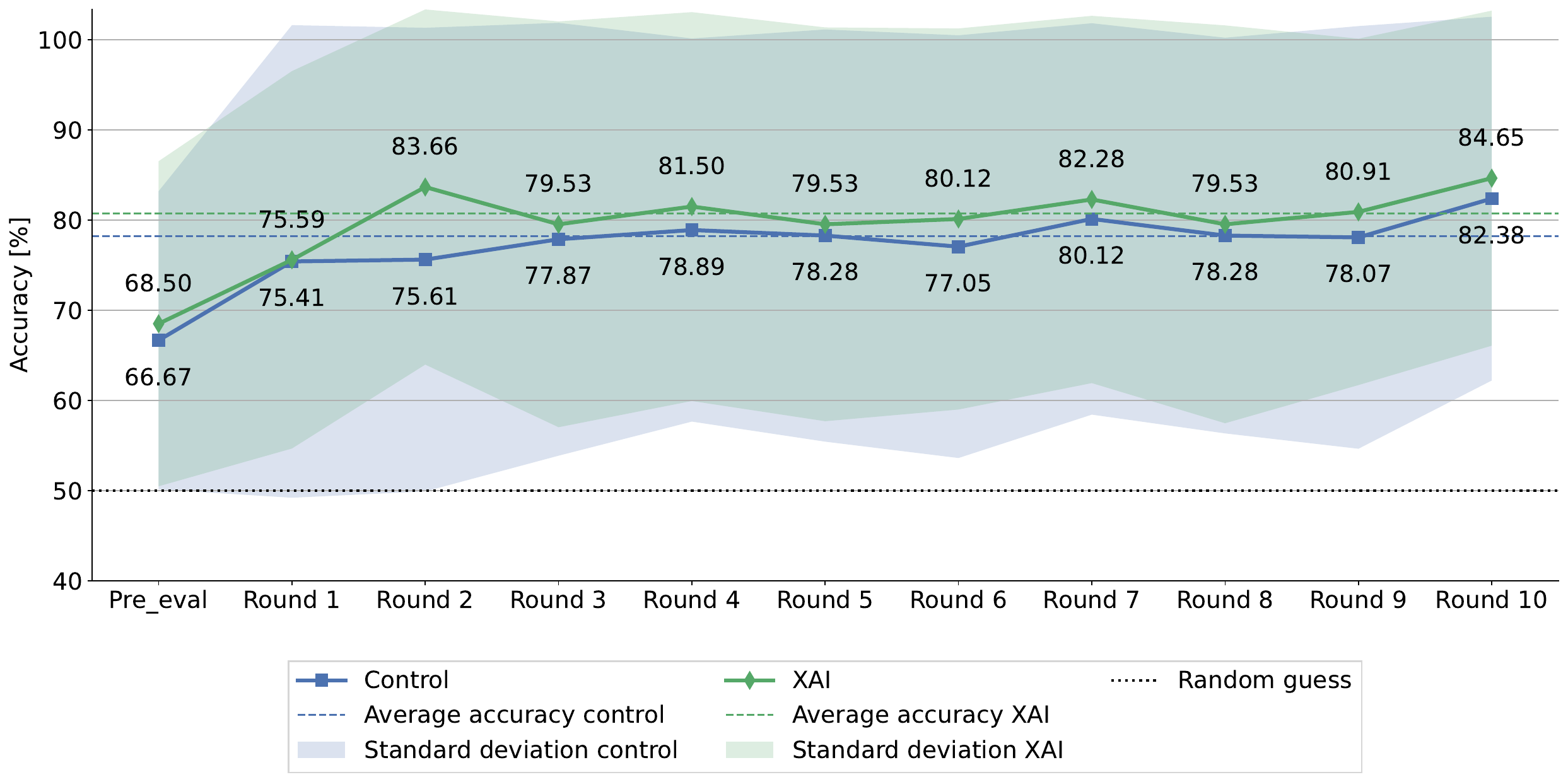}}
\captionsetup{justification=raggedright,singlelinecheck=false}
    \caption{Comparison of the learning curves in both conditions.}
    \label{results_image}       
\end{figure}

To answer \ref{RQ1}, we analyze participants' learning performance throughout the study. \Cref{results_image} shows the results over the course of the pre-evaluation and ten rounds. The plot highlights the average accuracies of participants in both conditions in each round associated with their standard deviation. Additionally, the plot shows the average accuracy in both conditions in dashed lines. We also draw the line for a random guess for this binary classification task.

The learning curve increases from the pre-evaluation round up to the second round in the XAI condition. In the control condition, it increases until round four. Throughout the next rounds of the experiment, the learning curve fluctuates in both conditions. Overall, the learning curve shows in both conditions its highest increase within the beginning of the study until round three. In the XAI condition, this increase is higher. This might be due to the effect that the additional visual explanations support novices' learning especially in the beginning of the study. In the last rounds, the learning curves climb again.Furthermore, the standard deviation decreases over the course of the ten rounds. We also compare the learning curves between both conditions. To do so we run a regression analysis in which we model the performance of each participant for one instance as dependent variable and the condition as binary independent variable. Additionally, to account for fixed effects of participants (as the same participants classifies four instances in each training round), we model the participant ID as random effect. The regression analysis reveals a positive significant effect for the independent variable ($p-value = .077$, $effect\;size = .143$).  This means that providing explanations generated by XAI leads to a higher performance throughout the rounds of the study. With the effect size of $.143$, we see that participants in the XAI condition have a higher performance and that XAI facilitates their learning. The mean of the XAI condition increases compared to the baseline condition. Thus, we confirm hypothesis \ref{hyp1}.  

Moreover, the accuracy (84.65 \%) in the last round of the XAI condition is higher than the accuracy (82.37\%) in the last round of the control condition. Additionally, in both conditions, the accuracy in the last round is significantly higher than the accuracy in the pre-evaluation (XAI condition: $p-value = 3.38\mathrm{e}^{-4}$, control condition: $p-value = 0.0212$).

\begin{figure}[htbp!]

    \centering{\includegraphics[width=0.9\textwidth]{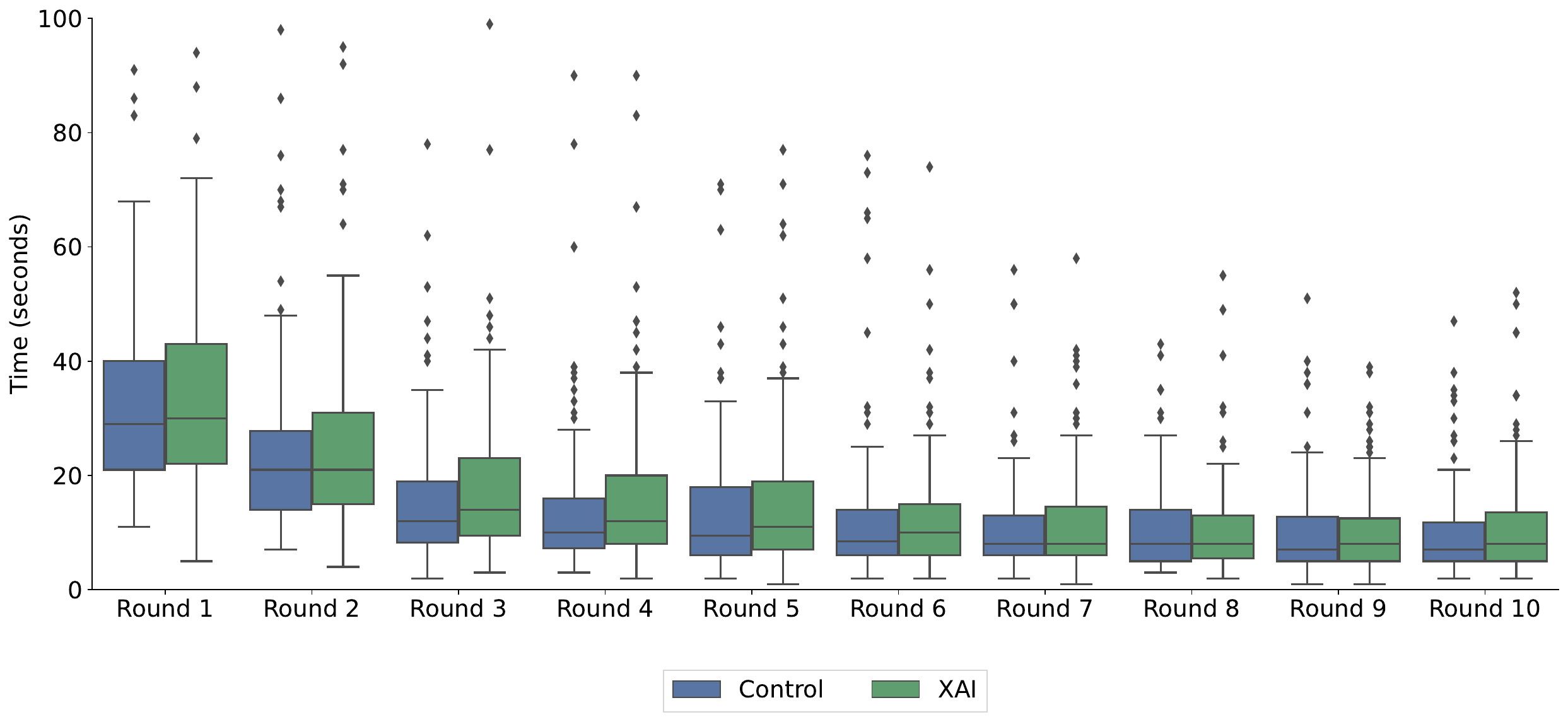}}
\captionsetup{justification=raggedright,singlelinecheck=false}
    \caption{Comparison of the learning times in both conditions.}
    \label{learningtimes}       

\end{figure}

In addition to the accuracy, we report the learning times for each condition. The learning times show a decrease throughout the rounds. The learning time in the first round is higher in the XAI condition compared to the control condition (see \Cref{learningtimes}). Similar to the performance, we conduct a regression analysis on the learning time. In this regression analysis we model the learning time as dependent variable, the treatment as binary independent variable and the participant ID as random effect. The regression analysis does not show a significant result ($p-value = .191$, $effect\;size = 4.191$). Hence, we do not find evidence to support hypothesis \ref{hyp2}.

\section{Impact of cognitive style in (X)AI-based learning systems}
\label{res_cogstyl}

In addition, we measure the visual cognitive style of the participants in this study through a post-task questionnaire to reveal the impact of these human factors on participants' learning performance and answer \ref{RQ2}. In the questionnaire, participants must answer 10 items on a five-point Likert scale. Accordingly, participants can achieve a maximum score of 50. We divide the participants into two groups: \textit{Visuals} and \textit{non-visuals}. We do this by calculating participants' scores on the questionnaire and distinguishing them into two groups by the median of scores on the questionnaire items. There are 72 participants with a visual cognitive style and 55 participants with a non-visual cognitive style in the XAI condition, and 64 and 58 in the control condition, respectively. 

The learning curve of participants with a visual cognitive style is above the one of the participants with a non-visual cognitive style for most of the rounds in both conditions. Moreover, in both conditions, both groups of participants (visuals and non-visuals) improve their accuracy in classifying the mammography images throughout the ten rounds. In the XAI condition, participants with a visual cognitive style achieve an average accuracy of $81.35\%$ while participants with a non-visual cognitive style achieve an accuracy of $79.90\%$ on average ($80.23\%$ and $75.95\%$ respectively in the control condition; XAI condition $p-value = 0.3407$, control condition $p-value = 0.045$). 

Overall, when comparing the means over all rounds between the two groups, participants with a visual cognitive style achieve a significantly higher performance than participants with a non-visual cognitive style ($p-value = 0.0134$). Thus we confirm hypothesis \ref{hyp3}.

The most surprising aspect of the data is in the increase of performance in the last round for the non-visual participants in the XAI condition. They end the study with a performance of $86.82\%$ and increase their performance from the pre-evaluation ($67.58\%$) by $19.24\%$-pts. Compared to the control condition, non-visual participants only increase their performance by $11.92\%$-pts.

Furthermore, there are no differences in learning time in the XAI conditions between participants with a visual and non-visual cognitive style. This might be due to the fact that both groups are provided with the same additional amount of information and, thus, take the same time to process the information compared to the control condition. In the control condition, however, the learning time of participants with a visual cognitive style is higher throughout the ten rounds. When comparing the average learning times of participants with the cognitive style in each condition, we find a significant difference between the learning times of participants with a visual cognitive style (higher) to participants with a non-visual cognitive style in the control condition ($p-value = 0.0273$). This can be explained as participants who identify with a visual cognitive style considering the provided visual information in more detail compared to participants who do not identify with this cognitive style. Thus, our data does not support hypothesis \ref{hyp4}.

In addition, we test whether novices' cognitive style moderates the effect of (X)AI-based learning on novices' learning performance and novices' learning time (see \Cref{research_model}). We do this to investigate whether participants' cognitive style impacts the relationship of providing explanations on the learning performance. To do so, we conduct several moderation analyses utilizing the process macro model of \citet{hayes2017introduction}. In these moderation analyses, we model the performance increase from the last round to the pre-evaluation round as the dependent variable, example-based learning (learning with XAI / without XAI) as the independent variable, and the visual cognitive style as the moderation variable. Thus, in this regression analysis, we explore whether there is an interaction effect between the independent variable and the moderation variable. In \Cref{mod_anal}, we see that there is a significant interaction effect which we highlight in bold.
An overview of the regression analyses is presented in \Cref{mod_anal}.

\begin{table}[htbp!]
\centering

\begin{threeparttable}

\begin{tabular}{m{3.5cm} R{1.5cm} R{1.5cm} R{0.002cm} R{1.5cm} R{1.5cm} } \hline
& \multicolumn{5}{c}{Dependent Variables} \\ \cmidrule{2-6}
& \multicolumn{2}{c}{Learning Performance} & & \multicolumn{2}{c}{Learning Time}
\\ \cmidrule{2-3} \cmidrule{5-6}
& coeff & se &  & coeff & se \\
\hline \hline
const   & .1193***   & .0350  && 23.3103 & 31.0910 \\
condition   & .0732   & .0502 && 64.7987 & 44.5649 \\
cognitive style   & .0722  & .0483 && -2.3572 & 42.9265  \\
interaction effect   & \textbf{-.1268*}  & .0679 && -60.8213 & 60.3386  \\\hline
$R$   & \multicolumn{2}{c}{0.1196} && \multicolumn{2}{c}{0.1142}  \\
$R^{2}$   & \multicolumn{2}{c}{0.0143} && \multicolumn{2}{c}{0.0130}  \\
$MSE$   & \multicolumn{2}{c}{0.0711} && \multicolumn{2}{c}{56065.8256}  \\
$F$   & \multicolumn{2}{c}{1.1857} && \multicolumn{2}{c}{1.0788}  \\\hline

\end{tabular}
    \begin{tablenotes}
    \centering
        \item[1]  \textit{p < 0.1} --- \textit{*}; \textit{p < 0.05} --- \textit{**}; \textit{p < 0.01} --- \textit{***}
    \end{tablenotes}
\label{mod_anal}
\end{threeparttable}
\captionsetup{justification=raggedright,singlelinecheck=false}
\caption{Results of the moderation analysis ($interaction effect = condition * cognitive style$).}
\end{table}

\begin{figure}[htbp!]

  \centering
  \includegraphics[width=0.6\textwidth]{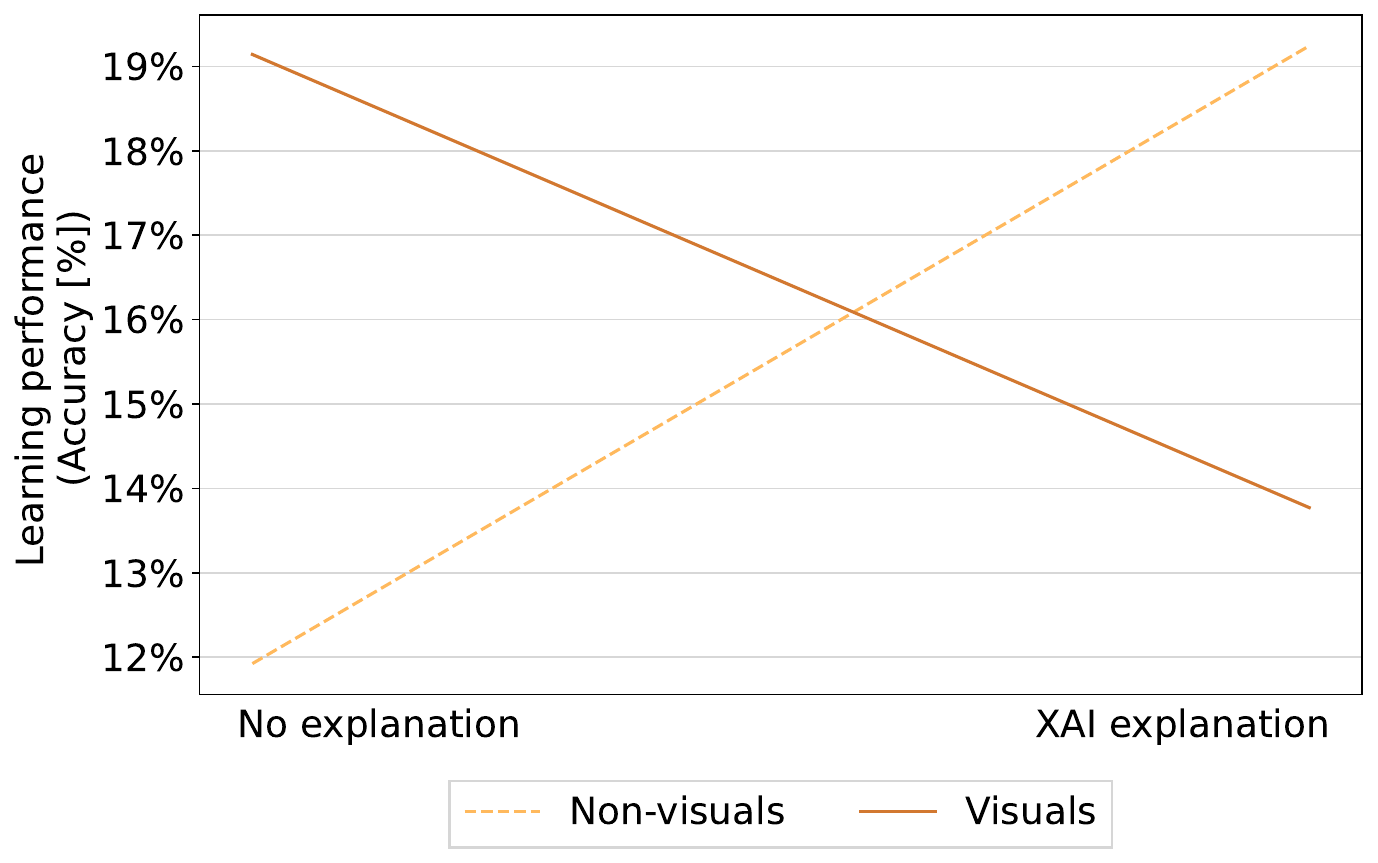}
\captionsetup{justification=raggedright,singlelinecheck=false}
    \caption{Interaction effect of visual cognitive style with example-based learning.}

\label{interaction_effects}

\end{figure}

The data shows a moderation effect of cognitive style on the relation of the example-based learning condition on novices' learning performance. This moderation effect is negative in its effect ($coeff=-.1268$); thus, for non-visual participants, the effect of explanations provided on the learning performance is higher. To account for potential randomization effects in building the subgroups of participants with different cognitive styles, we additionally conduct randomization checks by conducting a Monte Carlo cross-validation (size=.9; n=20). On average, the interaction effect has a size of $-.1284$ ($standard\;deviation\;effect\;size = .0177$, $average\;p-value = .0849$, $standard\;deviation p-value = .0422$). We also plot the interaction effect of visual cognitive style with the example-based learning condition in \Cref{interaction_effects}.

This data corroborates our previous findings as non-visual participants seem to benefit the most from the additional provided explanations. We can see in \Cref{interaction_effects} that their learning performance is higher when they are provided with this kind of support in the visual classification task. A possible explanation for this could be that non-visual participants cannot process the visual information in the task well. Thus, the additional explanations to indicate the relevant areas on the images facilitate their learning. 
There is no moderation effect of visual cognitive styles on novices' learning time.

\chapter{Discussion and Limitation}
\label{sec: discussion}

In this research paper, we present an (X)AI-based learning system for training novices to classify mammography images. Moreover, we test this (X)AI-based learning system in an online study through an experiment with between subjects design. We assess how explanations generated through XAI and novices' cognitive styles impact their learning performance.

\section{Implications for research and practice}

\textbf{Our findings show that XAI positively impacts novices' learning}. Overall, participants in both conditions increase their learning performance throughout the experiment. The learning performance in both conditions is significantly higher in the last round compared to their knowledge in the pre-evaluation round (XAI condition p-value = $2.29\mathrm{e}^{-11}$, control condition p-value = $2.20\mathrm{e}^{-10}$) which indicates that participants gain knowledge. In addition, the results of our experiment show that participants in the XAI condition perform better and learn more than participants in the control condition. Participants' learning curve in the XAI condition is, on average, significantly higher than the one of the control condition (see \Cref{results_image}). Similar to the study of \cite{schemmer2023towards}, who explore humans' learning in human-AI collaboration scenarios, we reveal the potential of utilizing XAI to facilitate novices' learning. With our results, we generate first insights into how (X)AI-based learning systems can facilitate novices' learning, representing a scalable alternative to conventional learning systems.

\textbf{Explanations facilitate non-visuals' learning in a visual task domain}. Moreover, we examine the effect of participants' cognitive style on their learning performance in the visual task of our experiment. On average, in both conditions, the learning performance of participants with a visual cognitive style is higher than that of participants with a non-visual cognitive style. The learning performance of participants in the XAI condition with a visual cognitive style is with $81.35\%$ highest through both conditions and cognitive style groups. In a previous study, \citet{riefle2022influence} reveal similar findings that participants' cognitive style impacts their understanding of explanations. Interestingly, the data shows that non-visual novices have the highest increase in learning performance. As non-visual participants might struggle more with finding patterns in the data to identify positive images, they receive the highest support through the explanations provided. With the aid of these explanations, they perform better than the non-visuals in the control condition. This interesting finding can be explained by the explanations provided to participants in the XAI condition. While non-visual participants over both conditions do not achieve the same final performance as visual participants, they might not be able to properly process the relevant visual information required for classifying the instances successfully (as prior research suggests, i.e., \cite{riding1991cognitive, kozhevnikov2007cognitive}). Thus, the indicated information (visual explanations) in the images (see \Cref{example}) aids their comprehension of relevant regions to classify the instances correctly.
By incorporating XAI and taking into account individual characteristics such as the cognitive styles of novices, organizations can improve the effectiveness of learning systems and can ultimately improve tasks at the workplace, e.g., for visual image classification like disease detection on X-rays or fault part detection of metal components.

\textbf{(X)AI-based learning systems present an effective addition to the toolset of knowledge managers}. Overall, the results show that (X)AI-based learning systems represent an effective learning approach for novices, as the learning performance increases in the XAI condition, while the learning time is similar to that in the control condition. In general, in the field of knowledge management, knowledge is distinguished into its \textit{explicit} and \textit{tacit} forms \citep{berry1987problem, polanyi2009tacit}. In \citet{nonaka1994dynamic}, the authors outline that tacit knowledge is difficult to transfer and can only be partially expressed. This poses a challenge for organizations to retain expert knowledge \citep{cavusgil2003tacit}. As \citet{vetter2023machine} show, machine learning can aid knowledge transfer in organizations. With our study, we show that XAI explanations can facilitate novices' learning by visually showing explicit AI knowledge that helps novices establish tacit knowledge of the underlying domain.
In addition, we take the first steps toward uncovering implications for the design of example-based learning systems by examining the learning curve of novices over the course of a multi-round study. The results show that novices increase their knowledge and learn to classify mammography images correctly. In a related study, \citet{kuhl2022human} examine the learning curve of humans and machines. They find that novices do not improve their learning performance after a certain number of samples due to cognitive overload. Our results show a similar effect, as the learning performance of novices in both conditions does not increase significantly after the third round. Thus, our findings can provide guidelines to designers to tailor such learning systems to the needs and preferences of individuals.
Especially for knowledge managers in organizations, it is crucial to find new ways to preserve the knowledge of experts and to transfer it to novices \citep{levy2011knowledge, burmeister2016knowledge}. 

\section{Limitations and outlook}
Our study certainly has some limitations. The images in our study include only mammography images with a clear positive or negative trend. Borderline cases were not included. This is reflected in the beginning of the learning curve in both conditions, which is above 50\%, representing the level of expertise of novices with no prior knowledge. Furthermore, the hardware setup used by the participants does not comply with the European guidelines for quality assurance in breast cancer screening and diagnosis \citep{perry2008european}, which may affect the difficulty of the task and create different prerequisites for the participants.
In addition, we only investigate the effect of visual explanations. So far, textual explanations have not been investigated as a factor on novices' learning performance. This highlights the need to explore this topic further with empirical research approaches. Moreover, with the current data of the study it is not possible to specify whether the effects on learning time are based on the additional provided information or on the specific explanation provided in this study. Future work could take first steps towards exploring these effects and look into factors that affect the learning time. Therefore, these limitations represent a fruitful starting point for future research on (X)AI-based learning systems.
In order to reveal possible implications not only for research but also for practice, the comparison of (X)AI-based learning systems with established systems in organizations is crucial. One kind of established learning systems in organizations are knowledge management systems \citep{levallet2018organizational}. Therefore, one future avenue of research is to investigate how the provision of information stored in such knowledge management systems affects the learning performance of novices. Furthermore, future research can explore comparing (X)AI-based learning systems to expert teaching.

\chapter{Conclusion}
\label{sec: conclusion}

This work presents a study design and empirical findings on (X)AI-based learning systems to improve novices' learning performance. To date, information system research lacks rigorous empirical studies on the impact of XAI on novices' learning and the influence of human factors. Therefore, through a between-subjects design study, we answer \ref{RQ1} how XAI facilitates novices' learning and empirically investigate how XAI can be used to teach novices a visual classification task. Our results demonstrate the effectiveness of these systems and are consistent with related work. They pave the way for research to evaluate the design of (X)AI-based learning systems more deeply. Furthermore, to address \ref{RQ2}, we conduct moderation analyses and show that their cognitive styles moderate their learning performance in this visual task. These findings can be used as a guide for designers of (X)AI-based learning systems.

Overall, we contribute to the ongoing discussion on knowledge retention and sharing between humans and AI. First, by investigating the use of (X)AI-based learning systems on novices' learning performance, we extend the existing discourse on novices' learning. Second, by elucidating novices' cognitive styles, we shed light on the impact of human factors in (X)AI-based learning processes. Extensive and rigorous research is needed to understand and utilize fully (X)AI-based learning systems in teaching novices. We invite researchers in the IS field to join this debate and hope to inspire scientists to participate actively in this endeavor.

\printbibliography

\end{document}